\documentclass{conm-p-l}
\usepackage{graphicx}

\usepackage{latexsym}
\usepackage{amsmath}
\usepackage{amssymb}
\usepackage{amsfonts}
\usepackage{graphicx}
\usepackage{epsfig}

\newcommand\arccot{\operatorname{arccot}}
\renewcommand\Im{\operatorname{Im}}

\begin{document}

\title{Experimental Mathematics and Mathematical Physics}

\author[]{David H.~Bailey}
\address{D.\,H.~Bailey:  Lawrence Berkeley National Laboratory, Berkeley, CA 94720}
\email{dhbailey@lbl.gov}

\author[]{Jonathan M.~Borwein}
\address{J.\,M.~Borwein: School of Mathematical and Physical Sciences, University of Newcastle, Callaghan, NSW 2308, Australia}
\email{jonathan.borwein@newcastle.edu.au}

\author[]{David Broadhurst}
\address{D.~Broadhurst:  Physics and Astronomy Department, Open University, Milton Keynes MK7 6AA, UK}
\email{D.Broadhurst@open.ac.uk}

\author[D.\,H.~Bailey, J.\,M.~Borwein, D.~Broadhurst, and W.~Zudilin]{Wadim Zudilin}
\address{W.~Zudilin:  School of Mathematical and Physical Sciences, University of Newcastle, Callaghan, NSW 2308, Australia}
\email{wadim.zudilin@newcastle.edu.au}

\thanks{D.\,H.~Bailey supported in part by the Director, Office of Computational and Technology Research, Division of Mathematical,
Information, and Computational Sciences of the U.S.\ Department of Energy, under contract no.~DE-AC02-05CH11231.
J.\,M.~Borwein supported in part by ARC}

\date{\today}

\begin{abstract}
One of the most effective techniques of experimental
mathematics is to compute mathematical entities such as integrals,
series or limits to high precision, then attempt to recognize the
resulting numerical values.  Recently these techniques have been
applied with great success to problems in mathematical physics.
Notable among these applications are the identification of some key
multi-dimensional integrals that arise in Ising theory, quantum
field theory and in magnetic spin theory.
\end{abstract}

\maketitle

\section{Introduction}

One of the most effective techniques of experimental mathematics is
to compute mathematical entities to high precision, then attempt to
recognize the resulting numerical values.  Techniques for
efficiently performing basic arithmetic operations and
transcendental functions to high precision have been known for
several decades, and within the past few years these have been
extended to definite integrals, sums of infinite series and limits
of sequences.  Recognition of the resulting numerical values is
typically done by calculating a list of $n$ possible terms on the
right-hand side of an identity, also to high precision, then
applying the {\sc pslq} algorithm \cite{cpslq,ppslq} to see if there is a
linear relation in this set of $n+1$ values.  If {\sc pslq} does find a
credible relation, then by solving this relation for the value in
question, one obtains a formula.  These techniques have been
described in detail in \cite{expmath1}, \cite{expmath2}, and
\cite{ema}.

In almost applications of this methodology, both in sophistication
and in computation time, the most demanding step is the computation
of the key value to sufficient precision to permit {\sc pslq} detection.
As we will show below, computation of some high-dimensional
integrals, for instance, often requires several hours on a highly
parallel computer system.  In contrast, applying {\sc pslq} to find a
relation among, say, 20 candidate terms, each computed to 500-digit
precision, usually can be done on a single-CPU system in less than a
minute.

In our studies of definite integrals, we have used either Gaussian
quadrature (in cases where the function is well behaved on a closed
interval) or the ``tanh-sinh'' quadrature scheme due to Takahasi and
Mori \cite{takahasi} (in cases where the function has an infinite
derivative or blow-up singularity at one or both endpoints).  For
many integrand functions, these schemes exhibit ``quadratic'' or
``exponential'' convergence -- dividing the integration interval in
half (or, equivalently, doubling the number of evaluation points)
approximately doubles the number of correct digits in the result.

The tanh-sinh scheme is based on the observation, rooted in the
\emph{Euler-Maclaurin summation} formula, that for certain
bell-shaped integrands (namely those where the function and all
higher derivatives rapidly approach zero at the endpoints of the
interval), a simple block-function or trapezoidal approximation to
the integral is remarkably accurate \cite[pg. 180]{atkinson1993}.
This principle is exploited in the tanh-sinh scheme by transforming
the integral of a given function $f(x)$ on a finite interval such as
$[-1, 1]$ to an integral on $(-\infty, \infty)$, by using the change
of variable $x = g(t)$, where $g(t) = \tanh (\pi/2 \cdot \sinh t)$.
The function $g(t)$ has the property that $g(x) \rightarrow 1$ as $x
\rightarrow \infty$ and $g(x) \rightarrow -1$ as $x \rightarrow
-\infty$, and also that $g'(x)$ and all higher derivatives rapidly
approach zero for large positive and negative arguments.  Thus one
can write, for $h > 0$,
\begin{eqnarray}\label{tanh-sinh}
\int_{-1}^1 f(x) \, {\rm d}x &=& \int_{-\infty}^\infty f(g(t)) g'(t)
\, {\rm d}t \; \approx \; h \sum_{j = -N}^N w_j f(x_j),
\end{eqnarray}
where the abscissas $x_j = g(hj)$, the weights $w_j = g'(hj)$, and
$N$ is chosen large enough that terms beyond $N$ (positive or
negative) are smaller than the ``epsilon'' of the numeric precision
being used.  In many cases, even where $f(x)$ has an infinite
derivative or an integrable singularity at one or both endpoints,
the transformed integrand $f(g(t)) g'(t)$ is a smooth bell-shaped
function for which the Euler-Maclaurin argument applies.  In these
cases, the error in this approximation (\ref{tanh-sinh}) decreases more rapidly than
any fixed power of $h$.  Full details are given in
\cite{quadrature}.

Both Gaussian quadrature and the tanh-sinh scheme are appropriate
for analytic functions on a finite interval.  Functions on a
semi-infinite intervals can be handled by a simple transformation
such as:
\begin{eqnarray*}
\int_0^\infty f(t) \, {\rm d}t &=& \int_0^1 f(t) \, {\rm d}t +
\int_0^1 \frac{f(1/t) \, {\rm d}t}{t^2}
\end{eqnarray*}
Oscillatory integrands such as $\int_0^\infty (1/x \sin x)^p \, {\rm
d}x$ can be efficiently computed by applying a clever technique
recently introduced by Ooura and Mori \cite{ooura}.  Let $x = g(t) =
M t / (1 - \exp (-2 \pi \sinh t))$.  Then in the case of $p = 2$,
for instance,
\begin{eqnarray*}
\int_0^\infty \left(\frac{\sin x}{x}\right)^2 \, {\rm d}x  &=&
\int_{-\infty}^\infty \left(\frac{\sin g(t)}{g(t)}\right)^2 \cdot
g'(t) \, {\rm d}t  \nonumber \\
  &\approx& h \sum_{k=-N}^N \left(\frac{\sin g(h k)}{g(h k)}\right)^2 \cdot g'(h k)
\end{eqnarray*}
Now note that if one chooses $M = \pi/h$, then for large $k$, the
$g(h k)$ values are all very close to $k \pi$, so the $\sin (g(hk))$
values are all very close to zero. Thus the sum can be truncated
after a modest number of terms, as in tanh-sinh quadrature.  In
practice, this scheme is very effective for oscillatory integrands
such as this.

In the next four sections we consider Ising integrals, Bessel moment integrals, `box'
integrals, and hyperbolic volumes arising from quantum field theory respectively.
We then conclude with a description of very recent work on multidimensional sums: Euler sums and MZVs.

\section{Ising integrals}

In a recent study, Bailey, Borwein and Richard Crandall applied
tanh-sinh quadrature, implemented using the ARPREC package, to study
the following classes of integrals  \cite{ising}. The $D_n$
integrals arise in the Ising theory of mathematical physics, and the
$C_n$ have tight connections to quantum field theory.
\begin{eqnarray*}
C_n &=& \frac4{n!} \int_{0}^{\infty} \cdots \int_{0}^{\infty}
\frac{1} { \left(\sum_{j=1}^{n} (u_j + 1/u_j) \right)^2} \frac{{\rm
d}u_1}{u_1}
\cdots \frac{{\rm d}u_n}{u_n}  \nonumber \\
D_n &=& \frac4{n!} \int_{0}^{\infty}  \cdots \int_{0}^{\infty}
\frac{\prod_{i < j} \left(\frac{u_i - u_j}{u_i + u_j}\right)^2} {
\left(\sum_{j=1}^{n} (u_j + 1/u_j) \right)^2} \frac{{\rm d}u_1}{u_1}
\cdots \frac{{\rm d}u_n}{u_n}  \nonumber \\
E_n &=& 2\int_0^1 \cdots \int_0^1 \left(\prod_{1\leq j < k \leq n}
   \frac{u_k - u_j}{u_k + u_j}\right)^2
{\rm d}t_2 \, {\rm d}t_3 \cdots dt_n,
\end{eqnarray*}
where (in the last line) $u_k = \prod_{i=1}^k t_i$.

Needless to say, evaluating these $n$-dimensional integrals to high
precision presents a daunting computational challenge.  Fortunately,
in the first case, we were able to show that the $C_n$ integrals can
be written as one-dimensional integrals:
\begin{eqnarray*}
C_n &=& \frac{2^n}{n!}  \int_0^{\infty} p K_0^n(p) \, {\rm d}p,
\end{eqnarray*}
where $K_0$ is the \emph{modified Bessel function} \cite{as}.  After
computing $C_n$ to 1000-digit accuracy for various $n$, we were able
to identify the first few instances of $C_n$ in terms of well-known
constants, e.g.,
\begin{eqnarray*}
C_3 &=& {\rm L}_{-3}(2) \; = \;  \sum_{n \geq 0} \left( \frac{1}{(3n + 1)^2} - \frac{1}{(3n + 2)^2} \right)  \nonumber \\
C_4 &=& \frac{7}{12} \zeta(3),
\end{eqnarray*}
where $\zeta$ denotes the Riemann zeta function.  When we computed
$C_n$ for fairly large $n$, for instance
\begin{eqnarray*}
C_{1024} &=&
0.63047350337438679612204019271087890435458707871273234\dots,
\end{eqnarray*}
we found that these values rather quickly approached a limit.  By
using the new edition of the \emph{Inverse Symbolic Calculator},
available at {\tt \small  http://ddrive.cs.dal.ca/\~{}isc}, this
numerical value can be identified as
\begin{eqnarray*}
\lim_{n \rightarrow \infty} C_n &=& 2 e^{-2 \gamma},
\end{eqnarray*}
where $\gamma$ is Euler's constant.  We later were able to prove
this fact---this is merely the first term of an asymptotic
expansion---and thus showed that the $C_n$ integrals are fundamental
in this context  \cite{ising}.

The integrals $D_n$ and $E_n$ are much more difficult to evaluate,
since they are not  reducible to one-dimensional integrals (as far
as we can tell), but with certain symmetry transformations and
symbolic integration we were able to reduce the dimension in each
case by one or two.  In the case of $D_5$ and $E_5$, the resulting
3-D integrals are extremely complicated, but we were nonetheless
able to numerically evaluate these to at least 240-digit precision
on a highly parallel computer system.  In this way, we produced the
following evaluations, all of which except the last we subsequently
were able to prove:
\begin{eqnarray*}
D_2 &=& 1/3  \nonumber \\
D_3 &=& 8 + 4\pi^2/3 - 27 \, {\rm L}_{-3}(2)  \nonumber \\
D_4 &=& 4 \pi^2/9 - 1/6 - 7 \zeta(3) / 2  \nonumber \\
E_2 &=& 6 - 8 \log 2  \nonumber \\
E_3 &=& 10 - 2 \pi^2 - 8 \log 2 + 32 \log^2 2  \nonumber \\
E_4 &=& 22 - 82 \zeta(3) - 24 \log 2 + 176 \log^2 2 - 256 (\log^3 2)/3 + 16 \pi^2 \log 2 - 22 \pi^2/3  \nonumber \\
E_5 &\stackrel{?}{=}& 42 - 1984 \, {\rm Li}_4(1/2) + 189 \pi^4/10
- 74 \zeta(3) - 1272 \zeta(3) \log 2 + 40 \pi^2 \log^2 2  \nonumber \\
&&  - 62 \pi^2/3 + 40 (\pi^2 \log 2)/3 + 88 \log^4 2 + 464 \log^2 2
- 40 \log 2,
\end{eqnarray*}
where ${\rm Li}$ denotes the polylogarithm function.  In the case of
$D_2, \, D_3$ and $D_4$, these are confirmations of known results.
We tried but failed to recognize $D_5$ in terms of similar constants
(the 500-digit numerical value is available if anyone wishes to
try).  The conjectured identity shown here for $E_5$ was confirmed
to 240-digit accuracy, which is 180 digits beyond the level that
could reasonably be ascribed to numerical round-off error; thus we
are quite confident in this result even though we do not have a
formal proof.

In a follow-on study \cite{meijer}, we examined the following
generalization of the $C_n$ integrals:
\begin{eqnarray*}
C_{n,k} &=& \frac4{n!} \, \int_{0}^{\infty}  \cdots
\int_{0}^{\infty}
 \frac{1} { \left(\sum_{j=1}^{n} (u_j + 1/u_j) \right)^{k+1}} \frac{{\rm d}u_1}{u_1} \cdots \frac{{\rm d}u_n}{u_n}.
\end{eqnarray*}
Here we made the initially surprising discovery---now proven in
\cite{BS} and in outline much earlier \cite{Bar}---that there are
linear relations in each of the rows of this array (considered as a
doubly-infinite rectangular matrix), e.g.,
\begin{eqnarray*}
0 &=& C_{3,0} - 84 C_{3,2} + 216 C_{3,4}  \nonumber \\
0 &=& 2 C_{3,1} - 69 C_{3,3} + 135 C_{3,5}  \nonumber \\
0 &=& C_{3,2} - 24 C_{3,4} + 40 C_{3,6}  \nonumber \\
0 &=& 32 C_{3,3} - 630 C_{3,5} + 945 C_{3,7}  \nonumber \\
0 &=& 125 C_{3,4} - 2172 C_{3,6} + 3024 C_{3,8}.
\end{eqnarray*}

\section{Bessel moment integrals}

In a more recent study of Bessel moment integrals, co-authored with
Larry Glasser \cite{besselmom}, the
first three authors were able to analytically recognize many of the
$C_{n,k}$ constants in the earlier study---because, remarkably,
these same integrals appear naturally in quantum field theory (for
odd $k$). We also discovered, and then proved with considerable
effort, that with $c_{n,k}$ normalized by $C_{n,k}={2^n}\, c_{n,k} /
(n!\,k!)$, we have
\begin{eqnarray*}
c_{3,0} &=& \frac{3 \Gamma^6(1/3)}{32 \pi 2^{2/3}}
  \; = \; \frac{\sqrt{3} \pi^3}{8} {}_3 F_2 \left(\begin{array}{c} 1/2, 1/2, 1/2 \\ 1, 1 \end{array} \Bigg| \frac 14 \right)  \nonumber \\
c_{3,2} &=& \frac{\sqrt{3} \pi^3}{288} {}_3 F_2 \left(\begin{array}{c} 1/2, 1/2, 1/2 \\ 2, 2 \end{array} \Bigg| \frac14 \right)  \nonumber \\
c_{4,0} &=& \frac{\pi^4}{4} \sum_{n=0}^\infty \frac{{2n \choose
n}^4}{4^{4n}}
      \; = \; \frac{\pi^4}{4} {}_4 F_3 \left(\begin{array}{c} 1/2, 1/2, 1/2, 1/2 \\ 1, 1, 1 \end{array} \Bigg| 1 \right)  \nonumber \\
c_{4,2} &=& \frac{\pi^4}{64} \left[4 \, {}_4 F_3 \left(\begin{array}{c} 1/2, 1/2, 1/2, 1/2 \\ 1, 1, 1 \end{array}\Bigg| 1 \right) \right. \nonumber \\
               &&  \left. - 3 \, {}_4 F_3 \left(\begin{array}{c} 1/2, 1/2, 1/2, 1/2 \\ 2, 1, 1 \end{array} \Bigg| 1 \right)\right] - \frac{3 \pi^2}{16},
\end{eqnarray*}
where ${}_p F_q$ denotes the \emph{generalized hypergeometric}
function \cite{as}. The  corresponding values for small odd second
indices are $c_{3,1} = 3 L_{-3}(2) / 4, \, c_{3,3} = L_{-3}(2) - 2 /
3, \, c_{4,1} = 7 \zeta(3) / 8$ and $c_{4,3} = 7 \zeta(3) / 32 - 3 /
16$.

Integrals in the Bessel moment study were quite challenging to
evaluate numerically.  As one example, we sought to numerically
verify the following identity that we had derived analytically:
\begin{eqnarray*}
c_{5,0} &=& \frac{\pi}{2} \int_{-\pi/2}^{\pi/2}
\int_{-\pi/2}^{\pi/2} \frac{{\bf K}(\sin\theta)\,{\bf K}(\sin\phi)}
{\sqrt{\cos^2\theta\cos^2\phi+4\sin^2(\theta+\phi)}} \,{\rm
d}\theta\,{\rm d}\phi\,,
\end{eqnarray*}
where ${\bf K}$ denotes the elliptic integral of the first kind
\cite{as}.  Note that this function has blow-up singularities on all
four sides of the region of integration, with particularly
troublesome singularities at $(\pi/2,-\pi/2)$ and $(-\pi/2, \pi/2)$
(see Figure 1).  Nonetheless, after making some minor substitutions,
we were able to evaluate (and confirm) this integral to 120-digit
accuracy (using 240-digit working precision) in a run of 43 minutes
on 1024 cores of the ``Franklin'' system at LBNL.

\begin{figure}
\begin{center}
\includegraphics[width=12cm]{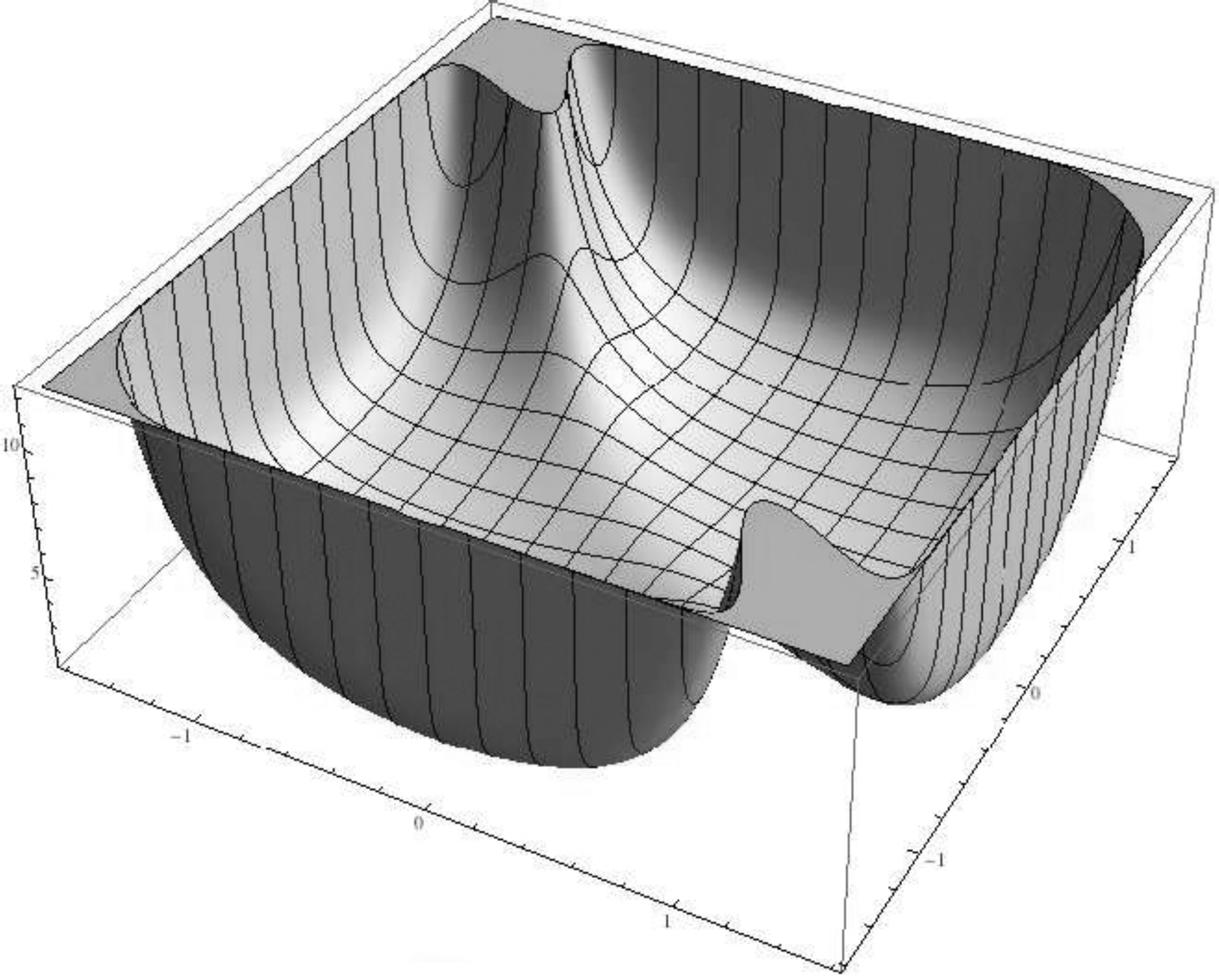}
\caption{Plot of $c_{5,0}$ integrand function}
\end{center}
\end{figure}

In a separate study, the first two authors studied correlation
integrals for the Heisenberg spin-1/2 antiferromagnet, as given by
Boos and Korepin, for a length-$n$ spin chain \cite[eqn.
2.2]{korepin3}:
\begin{eqnarray*}
P(n) &:=& \frac{\pi^{n(n+1)/2}}{(2 \pi i)^n} \cdot
\int_{-\infty}^\infty \int_{-\infty}^\infty \cdots
\int_{-\infty}^\infty
   U(x_1-i/2, x_2-i/2, \cdots, x_n-i/2)  \nonumber \\
   && \times \; T(x_1-i/2, x_2-i/2, \cdots, x_n-i/2) \, {\rm d} x_1 \, {\rm d} x_2 \cdots {\rm d} x_n
\end{eqnarray*}
where
\begin{eqnarray*}
U(x_1-i/2, x_2-i/2, \cdots, x_n-i/2) &=&  \frac{\prod_{1 \leq k < j
\leq n} \sinh [\pi(x_j - x_k)]}
   {\prod_{1 \leq j \leq n} i^n \cosh^n (\pi x_j)}  \nonumber \\
T(x_1-i/2, x_2-i/2, \cdots, x_n-i/2) &=& \frac{\prod_{1 \leq j \leq
n} (x_j-i/2)^{j-1} (x_j + i/2)^{n-j}}
   {\prod_{1 \leq k < j \leq n} (x_j - x_k - i)}. \hspace*{1em} \nonumber \\
\end{eqnarray*}

They computed numerical values for these $n$-fold integrals to as
great a precision as we could, then attempted to recognize them
using {\sc pslq}.  They found the following, which confirm some earlier
results obtained by others using  physical symmetry methods:
\begin{eqnarray*}
P(1) &=& \frac{1}{2}  \nonumber \\
P(2) &=& \frac{1}{3} - \frac{1}{3} \log 2  \nonumber \\
P(3) &=& \frac{1}{4} - \log 2 + \frac{3}{8} \zeta(3)  \nonumber \\
P(4) &=& \frac{1}{5} - 2 \log 2 + \frac{173}{60} \zeta(3)
 - \frac{11}{6} \zeta(3) \log 2 - \frac{51}{80} \zeta^2(3)
 - \frac{55}{24} \zeta(5) + \frac{85}{24} \zeta(5) \log 2  \nonumber \\
P(5) &=& \frac{1}{6} - \frac{10}{3} \log 2
 + \frac{281}{24} \zeta(3) - \frac{45}{2} \zeta(3) \log 2
 - \frac{489}{16} \zeta^2(3) - \frac{6775}{192} \zeta(5)  \nonumber \\
 && + \frac{1225}{6} \zeta(5) \log 2 - \frac{425}{64} \zeta(3) \zeta(5)
 - \frac{12125}{256} \zeta^2(5) + \frac{6223}{256} \zeta(7)  \nonumber \\
 && - \frac{11515}{64} \zeta(7) \log 2 + \frac{42777}{512} \zeta(3)
 \zeta(7)
\end{eqnarray*}

\begin{table}[ht]
\begin{center}
\begin{tabular}{|r|r|r|r|}
\hline
$n$ & Digits & Processors & Run Time \\
\hline
2 & 120 & 1 & 10 sec. \\
3 & 120 & 8 & 55 min. \\
4 & 60 & 64 & 27 min. \\
5 & 30 & 256 & 39 min. \\
6 & 6 & 256 & 59 hrs. \\
\hline
\end{tabular}
\vskip2mm
\caption{Run times and precision levels for spin integral
calculations}
\end{center}
\end{table}

These computations underscore the rapidly increasing cost of
computing integrals in higher dimensions.   Precision levels,
processor counts and run times are shown in Table~1.

\section{Box integrals}

Let us define \emph{box integrals} for dimension $n$ as
\begin{eqnarray*}
B_n(s) &:=& \int_0^1 \cdots \int_0^1  \left(r_1^2 + \dots + r_n^2\right)^{s/2} \, {\rm d}r_1 \cdots {\rm d}r_n \nonumber
\\
\Delta_n(s) &:=& \int_0^1 \cdots \int_0^1  \left((r_1 - q_1)^2 +
\dots +(r_n-q_n)^2\right)^{s/2} \,
  {\rm d} r_1 \cdots {\rm d}r_n \, {\rm d}q_1 \cdots {\rm d}q_n.
\end{eqnarray*}
As explained in previous treatments \cite{box1, box2}, these integrals have several physical interpretations:

\begin{enumerate}

\item $B_n(1)$ is the expected distance of a random point from the origin (or from any fixed vertex) of the $n$-cube.

\item $\Delta_n(1)$ is the expected distance {\it between} two random points of the $n$-cube.

\item $B_n(-n+2)$ is the expected electrostatic potential in an $n$-cube whose origin has a unit charge.  Such statements presume that
electrostatic potential in $n$ dimensions is $V(r) = 1/r^{n-2}$, and
say $\log r$ for $n = 2$; in other words, the negative powers of $r$
can also have physical meaning.

\item $\Delta_n(-n+2)$ is the expected electrostatic energy between two points in a uniform cube of charged ``jellium."

\item Recently integrals of this type have arisen in neuroscience ³ e.g., the average distance between synapses in a mouse brain.

\end{enumerate}

\noindent Note that the definitions show immediately that both
$\Delta_n(2m)$ and $B_n(2m)$ are rational when $m, n$ are natural
numbers. A pivotal, original treatment on box integrals is the 1976
work of Anderssen et al \cite{ander}.  There have been interesting
modern treatments of the $B_n$ and related integrals, as in
\cite{ten}, \cite[pg. 208]{expmath1}, \cite{weiss}, and
\cite{trott2005}.  Related material may also be found in
\cite{koepf, trott1998}.

Like the Ising integrals, some of these $n$-dimensional integrals
are reducible to 1-dimension integrals.  For instance, we found that
\begin{eqnarray*}
\Delta_3(-1) &=& \frac{2}{\sqrt{\pi}} \int_0^\infty \frac{(-1 + e^{-u^2} + \sqrt{\pi}\, u \,{\rm erf}(u))^3}{u^6} \, {\rm d}u.
\end{eqnarray*}
After calculating a 400-digit numerical value for this constant, we
were able to recognize it as
\begin{eqnarray*}
\Delta_3(-1) &=& \frac{1}{15} \left (6 + 6 \sqrt{2} - 12 \sqrt{3}
-10 \pi + 30 \log(1 + \sqrt{2}) + 30 \log(2 + \sqrt{3}) \right).
\end{eqnarray*}

\begin{table}
\begin{center}
\begin{tabular}{|c|c|c|}
\hline
$n$ & $s$ & $B_n(s)$  \\
\hline
any & even $s \geq 0$ & rational, e.g.: $B_2(2)=2/3$\\
1 & $s \neq -1$ & $\frac{1}{s+1}$\\[1.5pt]
\hline
2 & $-4$ & $-\frac14 - \frac{\pi}{8}$\\
2 & $-3$ & $-\sqrt 2$\\
2 & $-1$ & $2 \log(1 + \sqrt 2)$\\
2 & 1 & $\frac13 \sqrt 2 +  \frac13 \log(1 + \sqrt 2)$\\[1.5pt]
2 & 3 & $\frac{7}{20} \sqrt 2 +  \frac{3}{20} \log(1 + \sqrt 2)$\\[1.5pt]
2 & $s \neq -2$ & $\frac2{2+s} \,{}_2F_1\left(\frac{1}{2},-\frac{s}{2};\frac{3}{2};-1\right)$ \\[1.5pt]
\hline
3 & $-5$ & $- \frac16 \sqrt 3 - \frac1{12} \pi$ \\
3 & $-4$ & $- \frac32 \sqrt 2 \arctan \frac1{\sqrt 2}$ \\
3 & $-2$ & $- 3 G + \frac32 \pi \log(1 + \sqrt 2)  + 3\,{\rm Ti}_2(3 - 2 \sqrt 2)$ \\
3 & $-1$ & $- \frac14 \pi  + \frac32 \log \left(2 + \sqrt{3}\right)$ \\
3 & 1 & $ \frac14 \sqrt{3} -\frac1{24} \pi+\frac{1}{2} \log \left(2 + \sqrt{3}\right)$ \\[1.5pt]
3 & 3 & $ \frac25 \sqrt{3} -\frac1{60} \pi + \frac7{20} \log \left(2 + \sqrt{3}\right)$ \\[1.5pt]
\hline
\end{tabular}
\vskip2mm
\caption{Recent evaluations of Box integrals}
\end{center}
\end{table}

\begin{table}
\begin{center}
\begin{tabular}{|c|c|c|}
\hline
$n$ & $s$ & $B_n(s)$  \\
\hline
4 & $-5$ &  $-\sqrt{8}\,\arctan \left( \frac 1{\sqrt {8}} \right)$ \\
4 & $-3$ &  $4 \ G - 12\, {\rm Ti}_2(3 - 2 \sqrt 2)$ \\
4 & $-2$ & $\pi \,\log  \left( 2+\sqrt {3} \right) -2\,G-\frac{\pi^2}8$ \\
4 & $-1$ & $2\log 3-\frac23\,G+2\, {\rm Ti_2}\left(3-2\,\sqrt {2} \right)-\sqrt{8}\, \arctan\left(\frac 1{\sqrt{8}}\right)$ \\
4 & 1 &  $\frac{2}{5} - \frac G{10} + \frac{3}{10}\, {\rm Ti_2}\left(3-2 \sqrt{2}\right)+\log 3-\frac{7 \sqrt{2}}{10}\, \arctan\left(\frac1{\sqrt 8}\right)$ \\
\hline
   5 & $-3$ & ${\frac {110}{9}}\,G-10\,\log  \left( 2-\sqrt {3}
 \right)\, \theta-\frac1{8}\,{\pi }^{2}  $ \\[1.5pt]
 &&$-10\,{\rm Cl}_2 \left( \frac 13\,\theta+\frac 13\,\pi  \right)
 +10\,{\rm Cl}_2  \left( \frac 13\,\theta -\frac 16\,\pi \right) +\frac{10}{3}\,{\rm Cl}_2  \left( \theta+\frac 16\,\pi
 \right) $\\[1.5pt]
  && $+{\frac {20}{3}}\, {\rm Cl}_2\left( \theta+\frac 43\,\pi  \right) -{\frac {10}{3}}\,
{\rm Cl}_2 \left(\theta+\frac 53\,\pi\right)-{\frac {20}{3}}\,{\rm Cl}_2  \left(\theta+ {\frac {11}{6}}\,\pi \right)$\\[3pt]
5 & $-2$ & $\frac 83\,B_5(-6)-\frac13\,B_5(-4)+\frac 52\, \pi\log 3 +10\,{\rm Ti}_2 \left( \frac 13 \right) -10\,G $\\[2pt]
5 & $-1$ & $-{\frac {110}{27}}\,G+\frac{10}3\, \theta\log  \left( 2-\sqrt {3}
 \right)+\frac1{48}\,{\pi }^{2} $\\[1.5pt]
 &&$ +5\,\log  \left(\frac{ 1+\sqrt {5}}2 \right) -\frac 52\,\sqrt {3}\arctan \left(\frac 1{\sqrt {15}}
 \right)$\\[1.5pt]
  && $+\frac{10}3\,{\rm Cl}_2 \left( \frac 13\,\theta+\frac 13\,\pi  \right) -\frac{10}3\,{\rm Cl}_2  \left( \frac 13\,\theta -\frac 16\,\pi \right)
  -\frac{10}{9}\,{\rm Cl}_2  \left(
 \theta+\frac 16\,\pi
 \right)$\\[1.5pt]
  &&$ - {\frac {20}{9}}\, {\rm Cl}_2\left( \theta+\frac 43\,\pi  \right)  +{\frac {10}{9}}\,
{\rm Cl}_2 \left(\theta+\frac 53\,\pi\right)
  +{\frac {20}{9}}\,{\rm Cl}_2  \left(\theta+ {\frac {11}{6}}\,\pi \right)$\\[3pt]
5 & 1 &  $-\frac{77}{81}\,G+\frac {7}{9}\,\theta \log  \left( 2-\sqrt {3} \right)+\frac {1}{360}\,{\pi}^{2}+\frac16\,\sqrt {5} $\\[1.5pt]
&& $+\frac{10}3\,\log  \left(\frac{ 1+\sqrt {5}}{2 }\right)-\frac43\,\sqrt
{3}\arctan \left( \frac1{\sqrt {15}} \right)$\\[1.5pt]
&&$+ {\frac {7}{9}}\, {\rm Cl}_2\left(\frac 13\,\theta+\frac13\,\pi  \right)  -\frac {7}{9}
\, {\rm Cl}_2\left(\frac13\,\theta  -\frac16\,\pi \right) -{\frac {7}{27}}\, {\rm Cl}_2\left( \theta+\frac16\,\pi  \right)$\\[1.5pt]
 &&$   -{\frac {14}{27}}\,
{\rm Cl}_2\left( \theta+\frac 43\,\pi  \right)+{
\frac {7}{27}}\, {\rm Cl}_2\left(\theta+\frac 53\,\pi  \right)+{\frac {14}{27}}\,{\rm Cl}_2\left( {\theta+\frac {11}{6}}\,\pi  \right)$ \\[1.5pt]
\hline
\end{tabular}
\vskip2mm
\caption{Recent evaluations of Box integrals, continued; here $\theta:={\rm arctan}\left(\frac{16-3\sqrt{15}}{11}\right)$}
\end{center}
\end{table}

\begin{table}
\begin{center}
\begin{tabular}{|c|c|c|}
\hline
$n$ & $s$ & $\Delta_n(s)$  \\
\hline
2 & $-5$ & $\frac43 + \frac89 \sqrt 2$\\[2pt]
2 & $-1$ & $\frac43 - \frac43 \sqrt 2 + 4 \log(1 + \sqrt 2)$\\[2pt]
2 & 1 & $\frac2{15} + \frac{1}{15} \sqrt 2 + \frac13 \log(1 + \sqrt 2)$\\[2pt]
\hline
3 & $-7$ & $\frac{4}{5}-\frac{16 \sqrt{2}}{15}+\frac{2 \sqrt{3}}{5}+\frac{\pi }{15}$ \\[2pt]
3 & $-2$ & $2 \pi-12\ G + 12 \,{\rm Ti}_2\left(3-2 \sqrt{2}\right)+6 \pi  \log \left(1+\sqrt{2}\right)$ \\[1.5pt]
    && $+ 2\log2 -\frac52 \log 3-8 \sqrt{2} \arctan \left(\frac{1}{\sqrt{2}}\right) $\\[2pt]
3 & $-1$ & $\frac{2}{5}-\frac{2}{3} \pi+\frac{2}{5} \sqrt{2}-\frac{4}{5} \sqrt{3} +2 \log   \left(1+\sqrt{2}\right) +12 \log \left(\frac{1+\sqrt{3}}{\sqrt{2}}\right)-4 \log \left(2+\sqrt{3}\right)$ \\[2pt]
3 & 1 & $-{\frac {118}{21}}-\frac23\,\pi +{\frac {34}{21}}\,\sqrt {2}-\frac47\,\sqrt {3} +2\,\log  \left( 1+\sqrt {2} \right) +8\,\log  \left( \frac{ 1+\sqrt {3}}{\sqrt{2}}\right)$ \\[2pt]
3 & 3 & $-{\frac {1}{105}}-{\frac {2}{105}}\,\pi +{\frac {73}{840}}\,\sqrt {2}+\frac 1{35}\,\sqrt {3} +{\frac {3}{56}}\,\log  \left( 1+\sqrt {2} \right) +{\frac {13}{35}}\,\log  \left(\frac{ 1+\sqrt {3}}{\sqrt{2}} \right)$ \\[2pt]
\hline
\end{tabular}
\vskip2mm
\caption{Recent evaluations of Box integrals, continued}
\end{center}
\end{table}

\begin{table}
\begin{center}
\begin{tabular}{|c|c|c|}
\hline
$n$ & $s$ & $\Delta_n(s)$  \\
\hline
4& $-3$ & $-{\frac {128}{15}}+\frac{16}{3}\,\pi-8\,\log  \left( 1+\sqrt {2} \right)  -32\,\log  \left( 1+\sqrt {3} \right) +16
\,\log 2+20\,\log 3$ \\[1.5pt]
&&$ -\frac85\,\sqrt {2}+{\frac {32}{5}}\,\sqrt {3} -
32\,\sqrt {2}\arctan \left( \frac1{\sqrt {8}} \right)-96\,{\rm Ti}_2\left( 3-2\,\sqrt {2}
 \right) +32\,{G}$\\[3pt]
4& $-2$ & $-{\frac {16}{15}}\,\pi \,\sqrt {3}-\frac83\,\pi \,\log 2+
\frac{16}{3}\,\pi \,\log  \left( 1+\sqrt {3} \right) -\frac23\,{\pi }^{2}+\frac45\,\pi$\\[1.5pt]
&& $+\frac85\,\sqrt {2}\arctan \left( 2\,\sqrt {2} \right) +\frac25\,\log 3
-4\,\pi \,\log  \left( \sqrt {2}-1 \right)$ \\[1.5pt]
&& $+8{\rm Ti}_2\left( 3-2\,\sqrt {2} \right) -{\frac {40}{3}}G-\frac83\,\log 2$ \\[3pt]
4 & $-1$ & $-{\frac {152}{315}}-{\frac {8}{15}}\,\pi -{\frac {16}{5}}\,\log
2 +\frac25\,\log 3 +{\frac {68}{105}}\,
\sqrt {2}-{\frac {16}{35}}\,\sqrt {3}+\frac45\,\log  \left( 1+\sqrt {2}
 \right) $\\[1.5pt]
&& $ +{\frac {32}{5}}\,\log  \left( 1+\sqrt {3} \right) -\frac83\,G+8{\rm Ti}_2\left( 3-2\,\sqrt {2} \right) -\frac85\,\sqrt {2}\arctan \left( \sqrt {2}/4
 \right) $  \\[3pt]
 4 & 1 & $-{\frac {23}{135}}-{\frac {16}{315}}\,\pi -{\frac {52}{105}}\,\log 2
 +{\frac {197}{420}}\,\log 3 +{\frac {
73}{630}}\,\sqrt {2}+{\frac {8}{105}}\,\sqrt {3}$\\[1.5pt]
& & $+\frac1{14}\,\log  \left( 1+
\sqrt {2} \right) +{\frac {104}{105}}\,\log  \left( 1+\sqrt {3}
 \right) $ \\[1.5pt]
 && $-{\frac {68}{105}}\,\sqrt {2}\arctan
 \left(\frac{1}{\sqrt {8}} \right)-{\frac {4}{15}}\,{G}
 +\frac45\,{\rm Ti}_2\left( 3-2\,\sqrt {2} \right)$ \\[1.5pt]
\hline
5 & 1 &  $-{\frac {1279}{567}}\,{G}-{\frac {4}{189}}\,\pi +{\frac {4}{315}}\,
{\pi }^{2}-{\frac {449}{3465}}+{\frac {3239}{62370}}\,\sqrt {2}+{\frac {568}{3465}
}\,\sqrt {3}-{\frac {380}{6237}\,\sqrt {5}}$\\[1.5pt]
&&$+{\frac {295}{252}}\,\log 3 +{\frac {1}{54}}
\,\log  \left( 1+\sqrt {2} \right)+{\frac {20}{63}}\,\log  \left( 2+\sqrt {3} \right)+
{\frac {64}{189}}\,\log  \left(\frac{ 1+\sqrt {5}}2 \right)$\\[1.5pt]
&&$-{\frac {73}{63}}\,\sqrt {2}\arctan \left(\frac 1{\sqrt {8}} \right) -{\frac {8}{21}}\,\sqrt {3}\arctan
 \left(\frac1{\sqrt {15}} \right)  +{
\frac {104}{63}}\,\log  \left( 2-\sqrt {3} \right) \theta$\\[1.5pt]
&& $ +\frac57\,{\rm Ti}_2\left( 3-2\,\sqrt {2} \right)+{\frac
{104}{63}}\,{\rm Cl_2} \left( \frac13\,\theta+\frac13\,\pi  \right)
-{\frac {104}{63}}\,{\rm Cl_2} \left(\frac13\,\theta -\frac16\,\pi  \right)$\\[1.5pt]
&&$ -{\frac {104}{189}}\,{\rm Cl_2} \left(\theta+ \frac16\,\pi \right)
-{\frac {208}{189}}\,{\rm Cl_2} \left( \theta+\frac43\,\pi  \right)$ \\[1.5pt]
&& $+{\frac {104}{189}}\,{\rm Cl_2} \left(\theta+ \frac53\,\pi\right)
+{\frac {208}{189}}\,{\rm Cl}_{{2}} \left( \theta+{\frac {11}{6}}\,\pi  \right)$ \\[1.5pt]
\hline
\end{tabular}
\vskip2mm
\caption{Recent evaluations of Box integrals, continued}
\end{center}
\end{table}

A selection of results that we have found are shown in Tables 2, 3, 4 and~5.
\linebreak
Here $G$ denotes Catalan's constant, namely, $G := \sum_{n \ge
0} (-1)^n/(2n+1)^2$, $\theta =\linebreak \arctan\bigl((16 - 3 \sqrt{15})/11\bigr)$, Cl denotes
\emph{Clausen's function},
\begin{eqnarray*}
{\rm Cl}_2(\theta) &=& \sum_{n \geq 1} \frac{\sin(n\theta)}{n^2},
\end{eqnarray*}
 and Ti denotes Lewin's
\emph{inverse-tan function},
\begin{eqnarray*}
{\rm Ti_2}(x) &=&  \sum_{n \geq 0} (-1)^n \frac{x^{2n+1}}{ (2n+1)^2}.
\end{eqnarray*}

\section{Clausen functions and hyperbolic volumes}

In an unpublished 1998 study \cite{knots} two of the present authors (Borwein and Broadhurst) identified 998 closed hyperbolic 3-manifolds whose volumes are rationally related to Dedekind zeta values, with coprime integers $a$ and $b$ giving
\begin{eqnarray}
\label{vol}\frac{a}{b}\,{\rm vol}({\mathcal M}) &=& \frac{(-D)^{3/2}}{(2\pi)^{2n-4}}\,\frac{\zeta_K(2)}{2\zeta(2)}
\end{eqnarray}
for a manifold ${\mathcal M}$ whose invariant trace field $K$ has a single complex place, discriminant $D$, degree $n$, and Dedekind zeta value $\zeta_K(2)$. While the existence of integers $a, b$ can be established, via algebraic $K$-theory as in \cite{zagier3}, for the most part it was and is not possible to specify the rational $a/b$ other than empirically \cite{zagier3}.

The simplest identity implicit in (\ref{vol}) devolves to
\begin{eqnarray}
3\,{\rm Cl}_2(\alpha)-\,3{\rm Cl}_2(2\alpha)+{\rm Cl}_2(3\alpha) &=& \frac{7\sqrt{7}}{4}\, {\rm L}_{-7}(2),  \label{clausen}
\end{eqnarray}
with $\alpha=2 \arctan(\sqrt{7}),$ as is recorded in \cite[p. 91]{expmath1}. Here $L_{-7}(2):=\sum_{n>0} \left(\frac{n}{7}\right)/n^2$ is the primitive
Dirichlet $L$-series modulo~7 evaluated at~2 where $\left(\frac{n}{7}\right)$ is the Legendre symbol. This was rewritten in equivalent and more self-contained form as
\begin{eqnarray}
\label{arctan}
\frac{24}{7\sqrt{7}}\,\int _{\pi/3}^{\pi/2 }\, \log \left|
  \frac {\tan t + \sqrt{7}}{\tan t - \sqrt{7}} \right| \,{{\mathrm d}t}
&=& {\rm L}_{-7}(2)
\end{eqnarray}
in \cite[p. 61]{ema}---and elsewhere.

\begin{figure}
\centerline{\hss
\includegraphics[width=10cm]{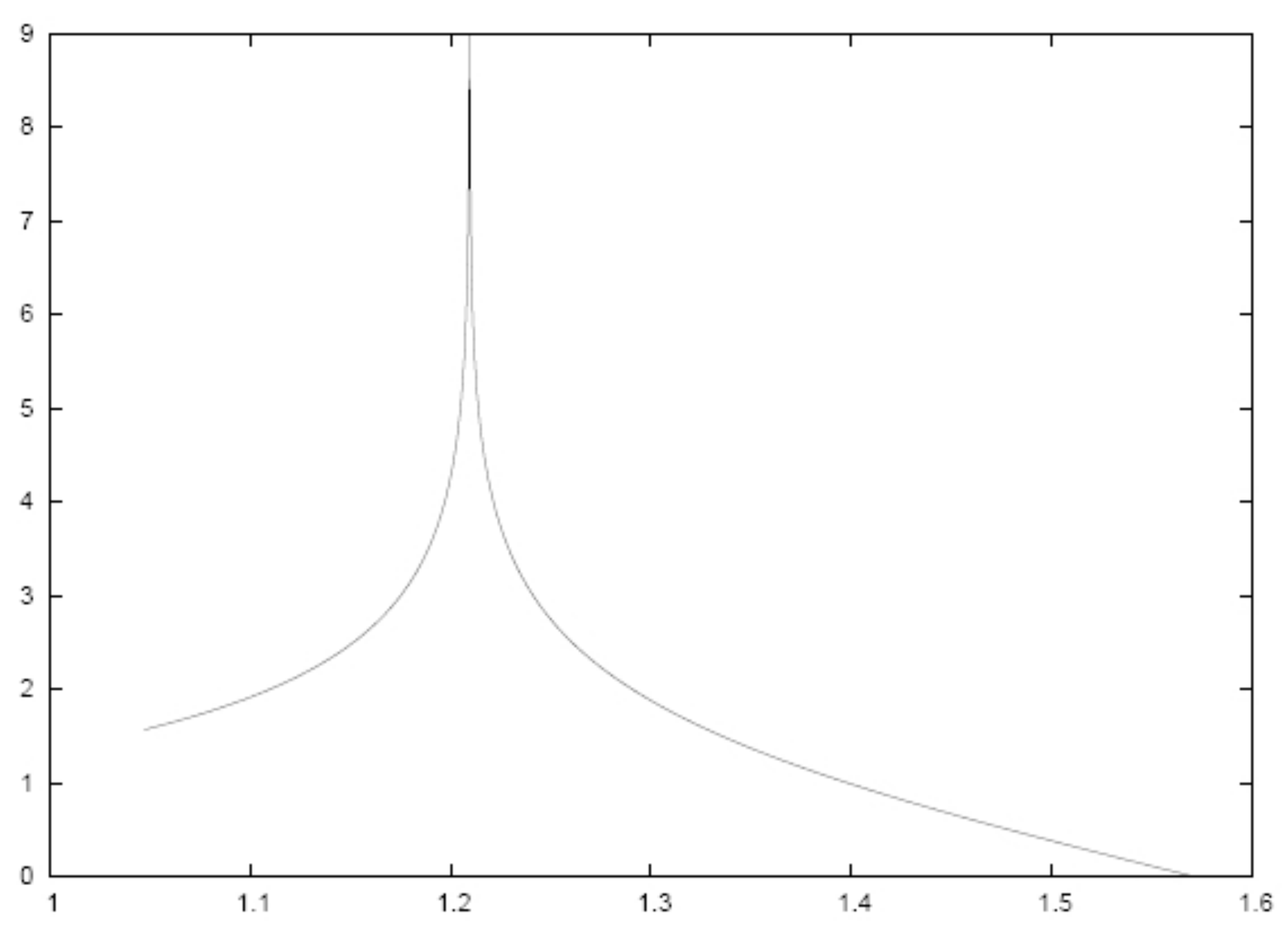}\hss}
\caption{Plot of integrand function in (\ref{arctan})}
\label{fig2}
\end{figure}

Note that the integrand function of (\ref{arctan}) has a nasty singularity at $\arctan(\sqrt{7})$
(see Figure~\ref{fig2}).  However, we were able to numerically evaluate this integral to
20,000-digit accuracy, by splitting the integral into two parts, namely on the intervals
$[\pi/3, \arctan(\sqrt{7})]$ and $[\arctan(\sqrt{7}), \pi/2]$.
Note that tanh-sinh quadrature can be used on each part, since it can readily handle
blow-up singularities at one or both endpoints of the interval of integration.
This run required 46 minutes on 1024 CPUs of the Virginia Tech Apple cluster.
The right-hand side was also evaluated, using \emph{Mathematica}, to 20,000-digit precision.
The two values agreed to $19,995$ digits \cite[pg. 61]{ema}.  Alternative representations
of the integral in~(\ref{arctan}) are given in~\cite{coffey}.

We shall now provide a proof of Eqn.~(\ref{clausen}) and hence of Eqn.~(\ref{arctan}).
Actually, an equivalent (if not obviously so) form of identity (\ref{clausen}), namely
\begin{eqnarray}
\label{zagier}
\zeta_{\mathbb Q(\sqrt{-7})}(2)
&=& \frac{\pi^2}{3\sqrt7}\Bigl\{A\Bigl(\cot\frac\pi7\Bigr)
+A\Bigl(\cot\frac{2\pi}7\Bigr)+A\Bigl(\cot\frac{4\pi}7\Bigr)\Bigr\}
\\
&=& \frac{2\pi^2}{7\sqrt7}\bigl\{2A(\sqrt7)+A(\sqrt7+2\sqrt3)
+A(\sqrt7-2\sqrt3)\bigr\}
\nonumber
\end{eqnarray}
with the notation
\begin{eqnarray*}
A(x)&:=&\int_0^x\frac1{1+t^2}\,\log\frac4{1+t^2}\,{\mathrm d}t
\; = \; {\rm Cl}_2(2\arccot x),
\end{eqnarray*}
is already established in \cite{zagier}.  The first equality in (\ref{zagier}) can be written as
\begin{eqnarray}\label{zagier1}
\zeta_{\mathbb Q(\sqrt{-7})}(2) &=& \frac{\pi^2}6\,{\rm L}_{-7}(2)
  \quad \; = \; \quad \zeta(2)\,{\rm L}_{-7}(2).
\end{eqnarray}
On noting that
\begin{eqnarray*}
\cot\arg\frac{\sqrt7+i}{2\sqrt2} &=& \sqrt7 \\
\cot\arg\frac{(1+i\sqrt7)(1\mp i\sqrt3)}{4\sqrt2} &=& \sqrt7\pm2\sqrt3 \\
{\rm Cl}_2(\theta) &=& \Im\sum_{n\ge1}\frac{e^{in\theta}}{n^2},
\end{eqnarray*}
we can translate the remaining, highly non-trivial, part of (\ref{zagier}) to
\begin{eqnarray*}
2A(\sqrt7)+A(\sqrt7+2\sqrt3)+A(\sqrt7-2\sqrt3) && \\
&& \hspace*{-15em} = \; 2\Im\sum_{n\ge1}\frac1{n^2}\biggl(\frac{\sqrt7+i}{2\sqrt2}\biggr)^{2n}
  +  \Im\sum_{n\ge1}\frac1{n^2}\biggl(\frac{(1+i\sqrt7)(1-i\sqrt3)}{4\sqrt2}\biggr)^{2n} \\
&& \hspace*{-15em} \quad + \Im\sum_{n\ge1}\frac1{n^2}\biggl(\frac{(1+i\sqrt7)(1+i\sqrt3)}{4\sqrt2}\biggr)^{2n}.
\end{eqnarray*}
Now we use
\begin{eqnarray*}
\biggl(\frac{(1+i\sqrt7)(1+i\sqrt3)}{4\sqrt2}\biggr)^2
&=& \mu e^{2\pi i/3} \\
\biggl(\frac{(1+i\sqrt7)(1-i\sqrt3)}{4\sqrt2}\biggr)^2
&=&\mu e^{-2\pi i/3} \\
\Im\biggl(\frac{\sqrt7+i}{2\sqrt2}\biggr)^{2n}
&=&\Im\biggl(\frac{3+i\sqrt7}4\biggr)^n
\; = \; -\Im(-\mu)^n \quad\text{for}\quad n=0,1,2,\dots,
\end{eqnarray*}
where $\mu:=(-3+i\sqrt7)/4$ has absolute value~1 and $\arg\mu=\alpha=2\arctan(\sqrt7)$, to write the latter equality as
\begin{eqnarray}
\label{zagier2}
2A(\sqrt7)+A(\sqrt7+2\sqrt3)+A(\sqrt7-2\sqrt3) && \\
&& \hspace*{-15em} = \; \Im\sum_{n\ge1}\frac{\mu^n(e^{2\pi in/3}+e^{-2\pi in/3}-2(-1)^n)}{n^2}
\nonumber \\
&& \hspace*{-15em} = \; \Im\biggl(\sum_{n\ge1}\frac{\mu^n}{n^2}
-\sum_{n\ge1}\frac{\mu^{2n}}{n^2}+\frac13\sum_{n\ge1}\frac{\mu^{3n}}{n^2}\biggr)
\nonumber \\
&& \hspace*{-15em} = \;  {\rm Cl}_2(\alpha)-{\rm Cl}_2(2\alpha)+\frac13{\rm Cl}_2(3\alpha),
\nonumber
\end{eqnarray}
where we  have applied the following two standard identities
\begin{eqnarray*}
\frac12\sum_{n\ge1}\frac{x^{2n}}{n^2}
  &=& \sum_{n\ge1}\frac{x^n}{n^2}+\sum_{n\ge1}\frac{(-1)^nx^n}{n^2} \\
\frac13\sum_{n\ge1}\frac{x^{3n}}{n^2}
  &=& \sum_{n\ge1}\frac{x^n}{n^2}+\sum_{n\ge1}\frac{e^{2\pi in/3}x^n}{n^2}
+\sum_{n\ge1}\frac{e^{-2\pi in/3}x^n}{n^2}
\end{eqnarray*}
for the dilogarithm function. It remains to substitute our finding (\ref{zagier2})
into (\ref{zagier}) and (\ref{zagier1}) to finish a proof of identity (\ref{clausen}).

The equivalent identity (\ref{arctan}) can be obtained by some reasonably straightforward but tedious manipulation of
the Clausen  integral representation
\begin{eqnarray}\label{arctan2}
{\rm Cl}_2(\theta)=- \int_0^\theta \log |2 \sin \sigma |\, {\rm d} \sigma
\end{eqnarray}
for $0 \le \theta <2\pi$, and an appropriate change of variables.

\medskip
As Don Zagier points out in~\cite{zagier}
\begin{quote}``\emph{we observe that the values of $A(x)$ at algebraic arguments satisfy
many non-trivial linear relations over the rational numbers; I know of
no direct proof, for instance, of the equality of the right-hand sides
of Eqns.~(5) and~(6).}''\end{quote}
Zagier's Eqns.~(5) and (6)   are  our identity (\ref{zagier}). Another result in~\cite{zagier},
Theorem~3, implies the identity
\begin{eqnarray}\label{zagier00}
\zeta_{\mathbb Q(\sqrt{-7})}(2)
&=& \frac{2\pi^2}{21\sqrt7}\biggl\{3A\biggl(\frac1{\sqrt7}\biggr)
  +3A\biggl(\frac3{\sqrt7}\biggr)+A\biggl(\frac5{\sqrt7}\biggr)\biggr\},
\end{eqnarray}
which may be thought of as complimentary to Eqn.~(\ref{zagier})
(see pg.~300 in~\cite{zagier} for details). Since
\begin{eqnarray*}
A\biggl(\frac1{\sqrt7}\biggr)
 &=& \operatorname{Cl}_2\bigl(2\arctan\sqrt7\bigr) \\
A\biggl(\frac3{\sqrt7}\biggr)
 &=& \operatorname{Cl}_2\biggl(2\arctan\frac{\sqrt7}3\biggr) \\
A\biggl(\frac5{\sqrt7}\biggr)
&=& \operatorname{Cl}_2\biggl(2\arctan\frac{\sqrt7}5\biggr),
\end{eqnarray*}
and
\begin{eqnarray*}
2\arctan\sqrt7=\alpha,
\quad
2\arctan\frac{\sqrt7}3=-2\alpha+2\pi,
\quad
2\arctan\frac{\sqrt7}5=3\alpha-2\pi,
\end{eqnarray*}
identity~(\ref{clausen}) follows from~(\ref{zagier00}) immediately.
Thus paper \cite{zagier} contains two different proofs of~(\ref{clausen})!

\medskip
Let us clarify the current status and somewhat-complicated history of various of the discoveries in \cite{knots}. Until recently the authors of \cite{knots} after discussion with Zagier believed  (\ref{zagier}) to be unproven. It was only when Zudilin spent time with Don Zagier in 2008 that he remembered his equivalent pre-dilogarithm (see \cite{zagier2,zagier3}) result in \cite{zagier}. Two of the present authors (Borwein and Broadhurst) \cite{knots} wrote
\begin{quote}``\emph{While the existence of such relations is
understood~\cite{zagier, zagier2, zagier3}, their precise forms
appear to be unpredictable, thus far, by deductive mathematics.
They are therefore ripe for the application of experimental mathematics.}''
\end{quote}

The great bulk of the results recorded in \cite{knots} remain unproven. They were discovered by intensive physically and mathematically motivated computation,
using \textsc{SnapPea}, \textsc{Pari-GP}, \textsc{Maple}, and other tools.

 Indeed, the cases $$D=-8, -11, -15, -20, -24$$ are challenging enough! These five respectively yield the following conjectured identities---each of which is open. First
\begin{eqnarray}
 27{\rm Cl}_2(\theta_2)-9{\rm Cl}_2(2\theta_2) +{\rm Cl}_2(3\theta_2)
&\stackrel{?}{=}& 8{\rm Cl}_2\left(\frac{2\pi}{8}\right) +8{\rm Cl}_2\left(\frac{6\pi}{8}\right),
\label{d8}
\end{eqnarray}
with $\theta_2:=2\arctan\sqrt2$. Secondly,
\begin{eqnarray}
15{\rm Cl}_2(\theta_{11})-10{\rm Cl}_2(2\theta_{11}) +{\rm Cl}_2(5\theta_{11})
&\stackrel{?}{=}&11\sum_{k=1}^5\left(\frac{k}{11}\right) {\rm Cl}_2\left(\frac{2\pi k}{11}\right),
\label{d11}
\end{eqnarray}
where $\theta_{11}:=2\arctan\sqrt{11}$ and $\left(\frac{k}{11}\right)$ is the Jacobi (or Legendre) symbol for the Dirichlet character.
Thirdly,
\begin{multline}
\label{d15}
24{\rm Cl}_2(\theta_{5,3})-12{\rm Cl}_2(2\theta_{5,3}) -8{\rm Cl}_2(3\theta_{5,3})
+6{\rm Cl}_2(4\theta_{5,3})
\\
\stackrel{?}{=}\quad 15\sum_{k=1}^7\left(\frac{k}{15}\right)
{\rm Cl}_2\left(\frac{2\pi k}{15}\right),
\end{multline}
with $\theta_{5,3}:=2\arctan\sqrt{5/3}$. Fourthly
\begin{multline}
\label{d20}
36{\rm Cl}_2(\theta_5)-30{\rm Cl}_2(2\theta_5) +4{\rm Cl}_2(3\theta_5)+3{\rm Cl}_2(4\theta_5)
\\
\stackrel{?}{=}\quad 20\sum_{k\in\{1,3,7,9\}}{\rm Cl}_2\left(\frac{2\pi k}{20}\right),
\end{multline}
with $\theta_5:=2\arctan\sqrt5$. Finally,
\begin{multline}
\label{d24}
60{\rm Cl}_2(\theta_{3,2})-18{\rm Cl}_2(2\theta_{3,2}) -4{\rm Cl}_2(3\theta_{3,2})+3{\rm Cl}_2(4\theta_{3,2})
\\
\stackrel{?}{=}\quad 24\sum_{k\in\{1,5,7,11\}}{\rm Cl}_2\left(\frac{2\pi k}{24}\right),
\end{multline}
with $\theta_{3,2}:=2\arctan\sqrt{3/2}$.
So, for the fifth time, we have a relation that is as easy to check numerically as it appears hard to derive. Needless to say, it would be interesting to check whether Zagier's 1986 theorems  in \cite{zagier} work for all such small values of~$D$; Theorem~2 in~\cite{zagier} looks sufficiently powerful for this task, while Theorem~3 therein depends critically on a delicate geometric construction and might be of use for $D=-11,-15,-20$. Moreover, is there a more transparent method to deduce identity~(\ref{clausen}) as well as
(\ref{d8})--(\ref{d24})?

As another example of the ubiquity of Clausen values, we complete this section with the most difficult integral evaluation required in \cite{box2}:
\begin{eqnarray}
\label{k1}
\qquad{\mathcal K}_1&:=&\int _{3}^{4}\!{\frac {{\rm arcsec} \left( x \right) }{\sqrt {{x}^{2}-
4\,x+3}}}\, {\rm d}x  \label{K1} \\
  &=& 3 {\rm Cl_2}  \left( \frac  \theta 3 \right) - \frac{3}{11} {\rm Cl_2} \left( \frac  \theta 3-\frac{\pi}{6} \right)
  - \frac{3}{11} {\rm Cl_2}  \left(\frac  \theta 3 + \frac{\pi}{6} \right)
  + \frac{18}{11} {\rm Cl_2}  \left( \theta -\frac{ \pi}3  \right) \nonumber \\
   && - \frac{15}{11} {\rm Cl_2} \left( \theta - \frac{2\pi}{3} \right)
    - \frac{3}{11} {\rm Cl_2} \left( \theta + \frac{\pi}{6} \right) - \frac{3}{11} {\rm Cl_2} \left( \theta - \frac{\pi}{6} \right) \nonumber\\
    && + \left( 2\,\theta - \frac {\pi}{2} \right) \log  \left( 2 - \sqrt{3} \right). \nonumber
\end{eqnarray}
Here
\begin{eqnarray*}
\theta &:=& \arctan\left(\frac{16-3\sqrt{15}}{11}\right).
\end{eqnarray*}
It may well be that this closed form (\ref{k1}) for ${\mathcal K}_1$ can be further simplified.

\section{Relations between MZVs and Euler sums}

We conclude with an application of experimental mathematics
to discover relations between
\emph{multiple zeta values} (MZVs) of
the form $$\zeta(s_1,s_2,\ldots,s_k)=\sum_{n_1>n_2>\ldots>n_k>0}
\frac{1}{n_1^{s_1}\ldots n_k^{s_k}}$$ with \emph{weight} $w=\sum_{i=1}^k
s_i$ and \emph{depth} $k$ and \emph{Euler sums} of the more general form
$$\sum_{n_1>n_2>\ldots>n_k>0}\frac{\epsilon_1^{n_1}\ldots\epsilon_k^{n_k}
}{n_1^{s_1}\ldots n_k^{s_k}}$$ with signs $\epsilon_i=\pm1$.
Both types of sum occur in evaluations of Feynman diagrams in
quantum field theory~\cite{BK,BBV} as mentioned in \cite{expmath1}. These sums are described in some mathematical detail in \cite[Chapter 3]{expmath2}.

First we recall the first Broadhurst--Kreimer conjectures (see \cite{BK} and also \cite{expmath2}) for the
enumeration of primitive MZVs and Euler sums of a given weight and depth.
Let $E_{n,k}$ be the number of independent Euler sums at weight $n>2$
and depth $k$ that cannot be reduced to primitive Euler sums of lesser depth
and their products. It is conjectured that~\cite{BK}
\begin{eqnarray*}
\prod_{n>2}\prod_{k>0}(1-x^n y^k)^{E_{n,k}}
&\stackrel{?}{=}& 1-\frac{x^3y}{(1-xy)(1-x^2)}. \label{conj1}
\end{eqnarray*}

We emphasise that, since the irrationality of odd values of
depth-one MZVs (i.e., Riemann's $\zeta$) is not settled, such
dimensionality conjectures are necessarily experimental. Now let
$D_{n,k}$ be the number of independent MZVs at weight $n>2$ and
depth $k$ that cannot be reduced to primitive MZVs of lesser depth
and their products. Thus we believe that $D_{12,4}=1$, since there
is no known relationship between the depth-4 sum
$\zeta(6,4,1,1)=\sum_{j>k>l>m} 1/(j^6k^4lm)$ and MZVs of lesser
depth or their products. It is conjectured that~\cite{BK}
\begin{eqnarray*}
\prod_{n>2}\prod_{k>0}(1-x^n y^k)^{D_{n,k}}
&\stackrel{?}{=}& 1-\frac{x^3y}{1-x^2}+\frac{x^{12}y^2(1-y^2)}{(1-x^4)(1-x^6)}.
\label{conj2}
\end{eqnarray*}

The final Broadhurst--Kreimer conjecture concerns the existence of relations
between MZVs and Euler sums of lesser depth. The now proven relation~\cite{BBV}
\begin{eqnarray*}
\zeta(6,4,1,1)&=&
\frac{64}{9}\zeta(\overline9,\overline3)
+\frac{371}{144}\zeta(9,3)
+3\zeta(2)\zeta(7,3)
+\frac{107}{24}\zeta(5)\zeta(7)\\
&& + \frac{1}{12}\zeta^4(3)
-\frac{3131}{144}\zeta(3)\zeta(9)
+\frac{7}{2}\zeta(2)\zeta^2(5)
+10\zeta(2)\zeta(3)\zeta(7)\\
&& + \zeta^2(2)\left[\frac{3}{5}\zeta(5,3)
-\frac{1}{5}\zeta(3)\zeta(5)
-\frac{18}{35}\zeta(2)\zeta^2(3)
-\frac{117713}{2627625}\zeta^4(2)\right]
\end{eqnarray*}
shows that the depth-4 MZV on the left can be expressed in terms
of Euler sums of lesser depth and their products. In fact, it suffices
to include the alternating double sum
$\zeta(\overline9,\overline3)=\sum_{j>k>0}(-1)^{j+k}/(j^9k^3)$,
where a bar above an argument of $\zeta$ serves to indicate
an alternating sign. In the language of~\cite{BK,BBV} this
is a ``\emph{pushdown}", at weight 12, of an MZV of depth 4 to an
Euler sum of depth 2.
Let $M_{n,k}$
be the number of primitive Euler sums of weight $n>2$ and depth $k$
whose products furnish a basis for all MZVs. It is conjectured that~\cite{BK}
\begin{eqnarray*}
\prod_{n>2}\prod_{k>0}(1-x^n y^k)^{M_{n,k}}
&\stackrel{?}{=}& 1-\frac{x^3y}{1-x^2}.
\label{conj3}
\end{eqnarray*}
Then by comparison of the output $D_{21,3}=6$, $D_{21,5}=9$, $D_{21,7}=1$
of~(\ref{conj2}) with the output $M_{21,3}=9$, $M_{21,5}=7$
of~(\ref{conj3}) we conclude that at weight 21, for example, three pushdowns
are expected from depth 5 to depth 3 and one from depth 7 to depth 5.

By massive use of the computer algebra language {\sc form},
to implement the shuffle algebras of MZVs and Euler sums,
the authors of~\cite{BBV} were recently able to reduce
all Euler sums with weight $w\le12$ and all MZVs with
$w\le22$ to concrete bases whose sizes are in precise
agreement with conjectures~(\ref{conj1},\ref{conj2}).
Moreover, further support to these conjectures came by studying
even greater weights, $w\le30$, using modular arithmetic.
However, such algebraic methods were insufficient
to investigate pushdown at weight 21. Instead the authors
resorted to a combination of the {\sc pslq} methods
reported in~\cite{ppslq} with the {\sc lll} algorithm~\cite{LLL} of
{\sc Pari-GP}~\cite{PARI}, finding empirical
forms for precisely the expected numbers of pushdowns at all weights
$w\le21$. Most notable of these is the pushdown from depth 7
to depth 5, at weight 21, in the empirical form
\begin{eqnarray*}
\label{push}
\zeta(6,2,3,3,5,1,1) &\stackrel{?}{=}&
-\frac{326}{81}\zeta(3,\overline6,3,\overline6,3)+\left\{\rm depth-5~MZV~products\right\}
\end{eqnarray*}
where the remaining 150 terms are formed by MZVs with depth
no greater than 5, and their products.

It is proven, by exhaustion, in~\cite{BBV} that the shuffle algebras do not allow the sum $\zeta(6,2,3,3,5,1,1)$ in equation (\ref{push}) to be reduced to MZVs of depth less than 7.  It is also proven that all other MZVs of weight 21 and depth 7 are reducible to $\zeta(6,2,3,3,5,1,1)$ and MZVs of depth less than 7.  Yet it appears to be far beyond the limits of current algebraic methods to prove that inclusion of the rather striking depth-5 alternating sum
\begin{eqnarray*}
\zeta(3,\overline6,3,\overline6,3)
&=& \sum_{j>k>l>m>n>0} \frac{(-1)^{k+m}}{(jk^2lm^2n)^3},
\end{eqnarray*}
with the rather simple coefficient $-326/81$, leaves the remainder reducible to MZVs of depth no greater than five.

Thus we are left with a notable empirical validation of a pushdown conjecture relevant to quantum field theory, crying out for elucidation.

\section{Conclusion}

We have presented here a brief survey of the rapidly expanding
applications of experimental mathematics (in particular, the
application of high-precision arithmetic) in mathematical physics.
It is worth noting that all but the penultimate of these examples have
arisen in the past five to ten years.  Efforts to analyze integrals
that arise in mathematical physics have underscored the need for
significantly faster schemes to produce high-precision values of
2-D, 3-D and higher-dimensional integrals.  Along this line, the
``sparse grid'' methodology has some promise \cite{smolyak, zenger}.

Current research is aimed at evaluating such techniques for
high-precision applications.  To illustrate the difficulty, we leave as a challenge to the reader the computation of the triple integral
\begin{eqnarray*}
\int_C \sqrt{f(u,v,w)-2}\, {\rm d}u \, {\rm d}v \, {\rm d}w  &=& 1.1871875\ldots,
\end{eqnarray*}
where $C:=[0,1/2]^3$ and
\begin{eqnarray*}
f(u,v,w) &:=& \cos ^2\left( (v+w)\pi  \right)+  \cos ^2\left( (u -v)\pi  \right) + \cos^2
 \left( (u+w)\pi  \right) \\
 && + \cos ^2\left( v\pi
 \right) + \cos^2 \left( u\pi  \right)  + \cos^2 \left( w\pi  \right)
\end{eqnarray*}
to, say, 32 decimal digit accuracy.

\raggedright

\end{document}